\documentclass[fleqn,10pt]{wlscirep}
\usepackage[utf8]{inputenc}
\usepackage[T1]{fontenc}
\usepackage{amsmath}
\newcommand{\lambdarm}[1]{$\lambda_{\mathrm{#1}}$}
\newcommand{\ploss}{$L_p$}
\newcommand{\sloss}{$L_s$}
\newcommand{\segnet}{$\phi_s$}

\newcommand{\newstuff}[1]{\textcolor{red}{#1}}

\title{Two-View Topogram-Based Anatomy-Guided CT Reconstruction for Prospective Risk Minimization}

\author[1,*]{Chang Liu}
\author[2,3]{Laura Klein}
\author[1,5]{Yixing Huang}
\author[2,3]{Edith Baader}
\author[4]{Michael Lell}
\author[2,3]{Marc Kachelrieß}
\author[2]{Andreas Maier}
\affil[1]{Pattern Recognition Lab, Friedrich-Alexander-Universität Erlangen-Nürnberg (FAU), Erlangen, Germany}
\affil[2]{Division of X-Ray Imaging and Computed Tomography, German Cancer Research Center  (DKFZ), Heidelberg, Germany}
\affil[3]{Department of Physics and Astronomy, Ruprecht-Karls-University Heidelberg, Heidelberg, Germany}
\affil[4]{Department of Radiology and Nuclear Medicine, Klinikum N\"urnberg, Paracelsus Medical University, N\"urnberg}
\affil[5]{Department of Radiation Oncology, Universitätsklinikum Erlangen, Friedrich-Alexander-Universität Erlangen-Nürnberg (FAU), Erlangen, Germany}

\affil[*]{chang.ch.liu@fau.de}

\begin{abstract}
To facilitate a prospective estimation of CT effective dose and risk minimization process, a prospective spatial dose estimation and the known anatomical structures are expected. To this end, a CT reconstruction method is required to reconstruct CT volumes from as few projections as possible, i.e. by using the topograms, with anatomical structures as correct as possible.
In this work, an optimized CT reconstruction model based on a generative adversarial network (GAN) is proposed. The GAN is trained to reconstruct 3D volumes from an anterior-posterior and a lateral CT projection. To enhance anatomical structures, a pre-trained organ segmentation network and the 3D perceptual loss are applied during the training phase, so that the model can then generate both organ-enhanced CT volume and the organ segmentation mask. 
The proposed method can reconstruct CT volumes with PSNR of 26.49, RMSE of 196.17, and SSIM of 0.64, compared to 26.21, 201.55 and 0.63 using the baseline method. 
In terms of the anatomical structure, the proposed method effectively enhances the organ shape and boundary and allows for a straight-forward identification of the relevant anatomical structures. 
We note that conventional reconstruction metrics fail to indicate the enhancement of anatomical structures.
In addition to such metrics, the evaluation is expanded with assessing the organ segmentation performance. The average organ dice of the proposed method is 0.71 compared with 0.63 in baseline model, indicating the enhancement of anatomical structures.  

\end{abstract}
\begin{document}

\flushbottom
\maketitle
\thispagestyle{empty}

% \noindent Please note: Abbreviations should be introduced at the first mention in the main text – no abbreviations lists. Suggested structure of main text (not enforced) is provided below.

\section*{Introduction}

Computed tomography (CT) imaging provides non-invasive insights into the human body with a high image quality and only short acquisition time compared to other modalities. Therefore, CT imaging has become an integral part of clinical routine and research. However, in order to reconstruct CT volumes with a diagnostic image quality, a sufficient number of measured projections must be acquired which inevitably exposes the patient to ionizing radiation, i.e., X-rays. Therefore, dose reduction is an important research topic in CT imaging. There are different methods to achieve dose reduction, both hardware- and software-based. These methods include but are not limited to the usage of pre-filters, iterative reconstruction algorithms, and dose-shielding methods. One other method that is routinely used is to adjust the tube current of the X-ray source depending on the angular position $\alpha$ of the X-ray source and the z-position, so called tube current modulation (TCM) \cite{gies1999dose, kalender1999dose}. More precisely, TCM methods aim at minimizing the mAs-product by adapting the tube current as a function of attenuation for a given view. The attenuation can for example be estimated based on the topogram acquired prior to the CT scan. 

However, the mAs-product is only a surrogate parameter for actual patient dose, since some organs are more sensitive to the radiation than others. It would be of advantage to also account for these sensitivities in the tube current optimization. Thereby, the effective dose $D_{\text{eff}}$ is defined as the sum of the dose absorbed by the organ-at-risks (OAR) during the exposure, weighted with the organ-specific tissue weighting factor. The tissue weighting factors correspond to the radiation sensitivity of the individual organs and structures are provided by the international commission on radiological protection (ICRP) \cite{Valentin2007-oi, Icrp2007-yx}. The factors also reflect the risk of radiation induced for cancer. To this end, it is necessary to be able to estimate organ doses, namely effective dose, prior to the CT scan and then optimize for $D_{\text{eff}}$. Recently, a risk-minimizing tube current modulation (riskTCM) has been proposed that requires a dose distribution and organ segmentation as input parameters \cite{riskTCM}. In particular, this method assumes an initial coarse CT reconstruction and the voxel-wise segmentation of all relevant organs. Given the known sensitivities with respect to ionizing radiation of these organs, the effective dose is estimated on a per-view basis. Usually, dose estimation is performed using Monte Carlo methods. Such methods, however, are very time consuming and would prohibit an application of riskTCM in clinical practice. Hence, spatial dose distribution is estimated in quasi-real-time from a given CT volume using a deep neural network proposed by Maier et al. \cite{Maier2022}. Organ dose is then obtained using known organ segmentation. With the effective dose for each potential view in the desired scan range, a tube current curve is then computed that allows maintaining diagnostic image quality while minimizing the patient risk.

\begin{figure}[ht]
\centering
\includegraphics[width=0.8\linewidth]{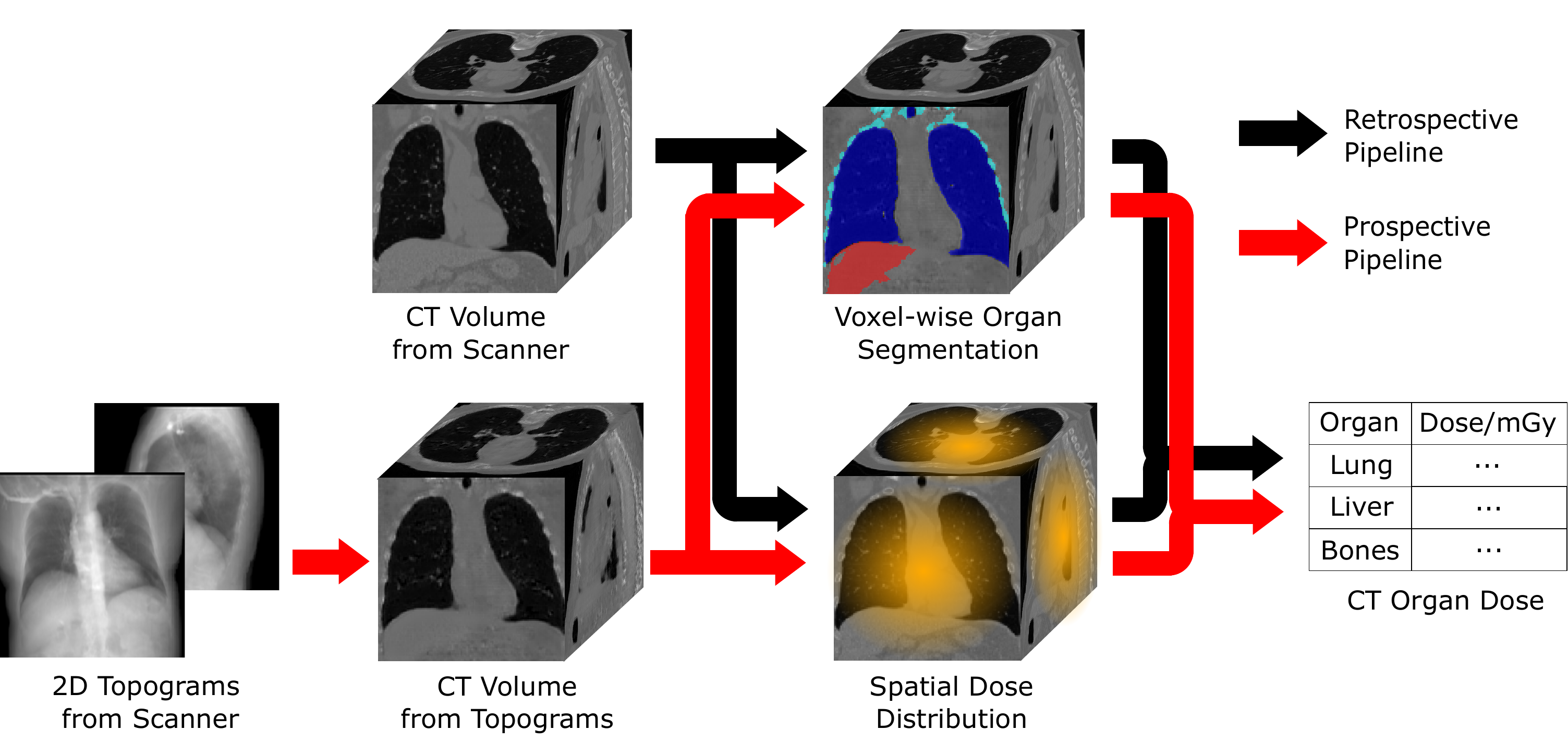}
\caption{
Illustration of prospective and retrospective CT organ dose estimation pipelines. Many existing methods to estimate CT organ dose are designed as retrospective pipelines, which can only be applied after the scanning. For application like CT risk minimization, a prospective pipeline for organ dose estimation is required.
}  %note label inside caption
\label{fig_intro_pipeline} 
\end{figure}

% To achieve this, a method that estimates a coarse CT reconstruction before the scanning is needed for prospective CT organ dose estimation. As shown in Fig. \ref{fig_intro_pipeline}, only few projections are required to estimate CT organ dose and thus it is more suitable for CT risk minimization. In order to avoid additional X-ray projections, we refactor the research problem to the reconstruction of a coarse CT volume from only two orthogonal topograms, referred to as X-ray projections in the following manuscript, while enabling dose estimation and organ segmentation. 
To achieve this, a method that estimates a coarse CT reconstruction before the scanning is needed. 
As shown in Figure \ref{fig_intro_pipeline}, starting from only few projections provides a reasonable pipeline to facilitate the CT risk optimization rather than the retrospective CT dose estimation.  
% only few projections are required to estimate CT organ dose and thus it is more suitable for CT risk minimization. 
In order to avoid additional X-ray projections, we refactor the research problem to the reconstruction of a coarse CT volume from only two orthogonal topograms, referred to as X-ray projections in the following manuscript.

With the emerge of deep learning (DL)-based medical image processing methods, some generative adversarial network (GAN) methods have been established related with CT reconstruction from only few views \cite{goodfellow2014generative}. Ying et al. proposed X2CT-GAN\cite{ying2019x2ct} that performs a domain transfer task from X-ray projections to CT volumes, where a network for effective 2D-to-3D image generation is proposed. The authors also address the superiority of using two X-ray projections, i.e.\ from anterior-posterior (a.p.) and lateral (lat.) direction, compared to only a single view. On top of the X2CT-GAN, Ling et al. proposed a conditional variational autoencoder (cVAE)-based GAN\cite{LingCVAE} to enhance the regularization of the generator. Ratul et al. improved the generator with additional input of the organ segmentation of the X-ray projections from a.p. direction\cite{RatulCCXray}. Montoya et al. proposed ScoutCT-Net that first backprojects the topograms into an initial CT volume, and refine the initial volume using another network \cite{Montoya2022-nh}. 
Similarly, most proposed methods aim to improve the CT reconstructions by voxel-wise metrics, while the anatomical information, such as the shape and location of organs and structures, are usually ignored.

In this work, we propose an anatomy-guided GAN for CT reconstruction from only two X-ray projections which can facilitate the implementation of the risk-specific TCM methods. 
More specifically, a 3D perceptual loss \ploss and a 3D segmentation loss \sloss are implemented into the overall loss function for training the GAN, leading to a loss function that also optimize for better anatomical information: 
\begin{equation}
    L_G' = L_G + \lambda_p L_p + \lambda_s L_s,
\end{equation}
where $L_G$ is the original generator loss, focusing on voxel-wise similarity, and $\lambda_p$ and $\lambda_s$ are constants that control the enhancement. We demonstrate that the combined use of \ploss and \sloss can lead to the enhancement of anatomical structures in the reconstructed volumes. The implementation of \ploss and \sloss will be in detail explained in following sections. Our proposed method enhances the organ shape and boundary during the training phase and thus will not increase the computational complexity during inference time. 

% \begin{figure}
% \centering
% \includegraphics[width=0.5\linewidth]{Figure_2.pdf}
% \caption{The illustration of the proposed method. the reconstruction GAN is trained, regularized by the proposed 3D segmentation loss and 3D perceptual loss. During inference, the CT generator in the GAN reconstructs the CT volume with input of two X-ray projections, and the S can be used to provide the organ segmentation.
% \label{fig_illu}}
% \end{figure}

\section*{Results}

% Up to three levels of \textbf{subheading} are permitted. Subheadings should not be numbered.

% \subsection*{Subsection}

% Example text under a subsection. Bulleted lists may be used where appropriate, e.g.

% \begin{itemize}
% \item First item
% \item Second item
% \end{itemize}

% \subsubsection*{Third-level section}
 
% Topical subheadings are allowed.

\begin{figure}[ht]
  \centering
  \includegraphics[width=16cm]{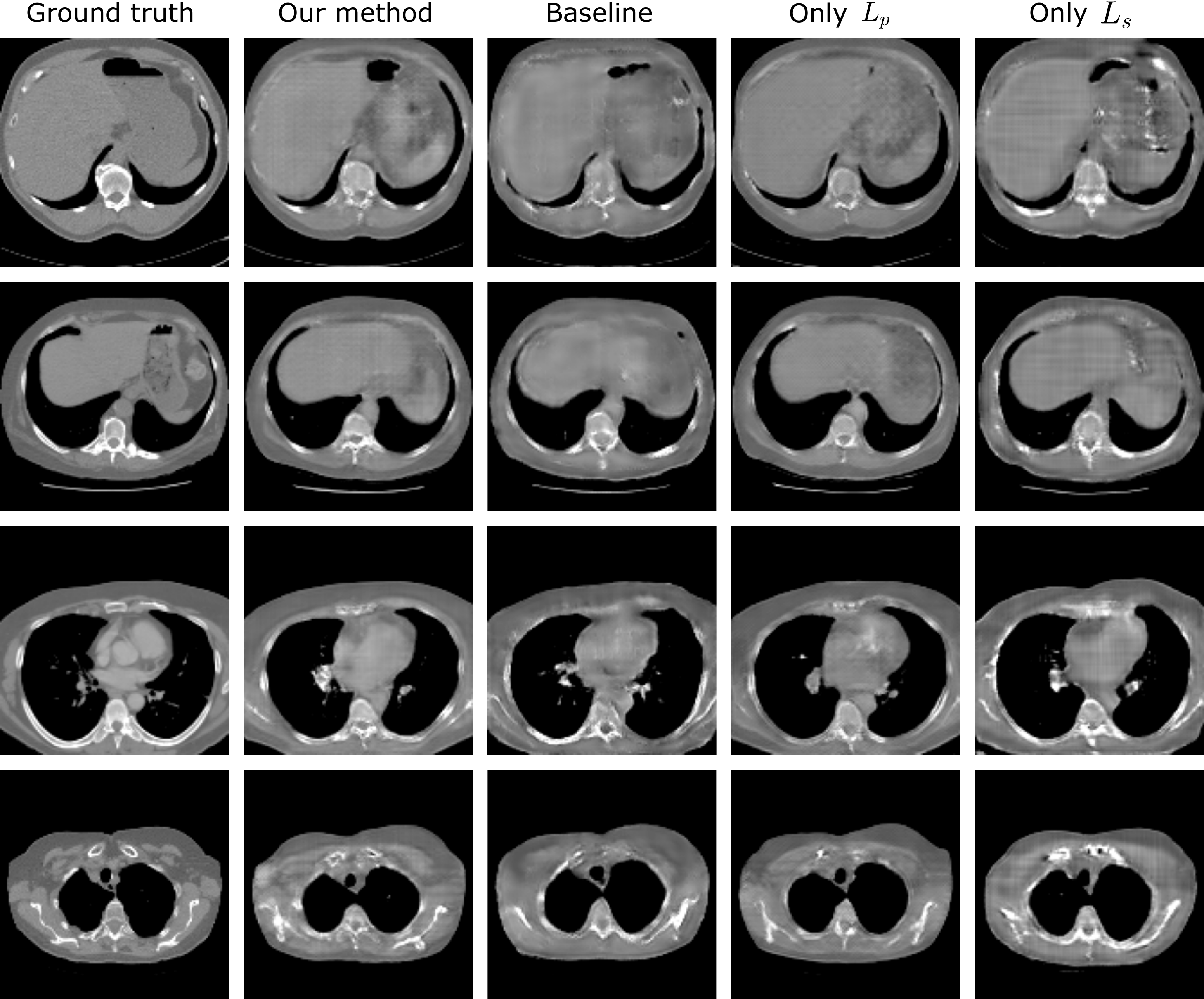}
  \caption{Exemplary slices of the reconstructed CT volumes using the proposed and baseline method, in comparison with the ground truth. 
  % In the figure, '\emph{P} Loss' indicates 3D perceptual loss, '\emph{S} Loss' indicates 3D segmentation loss. The proposed method applies both losses.
   }  %note label inside caption
  \label{fig_res_example} 
\end{figure}

\begin{figure}[ht]
    \centering
        \includegraphics[width=16cm]{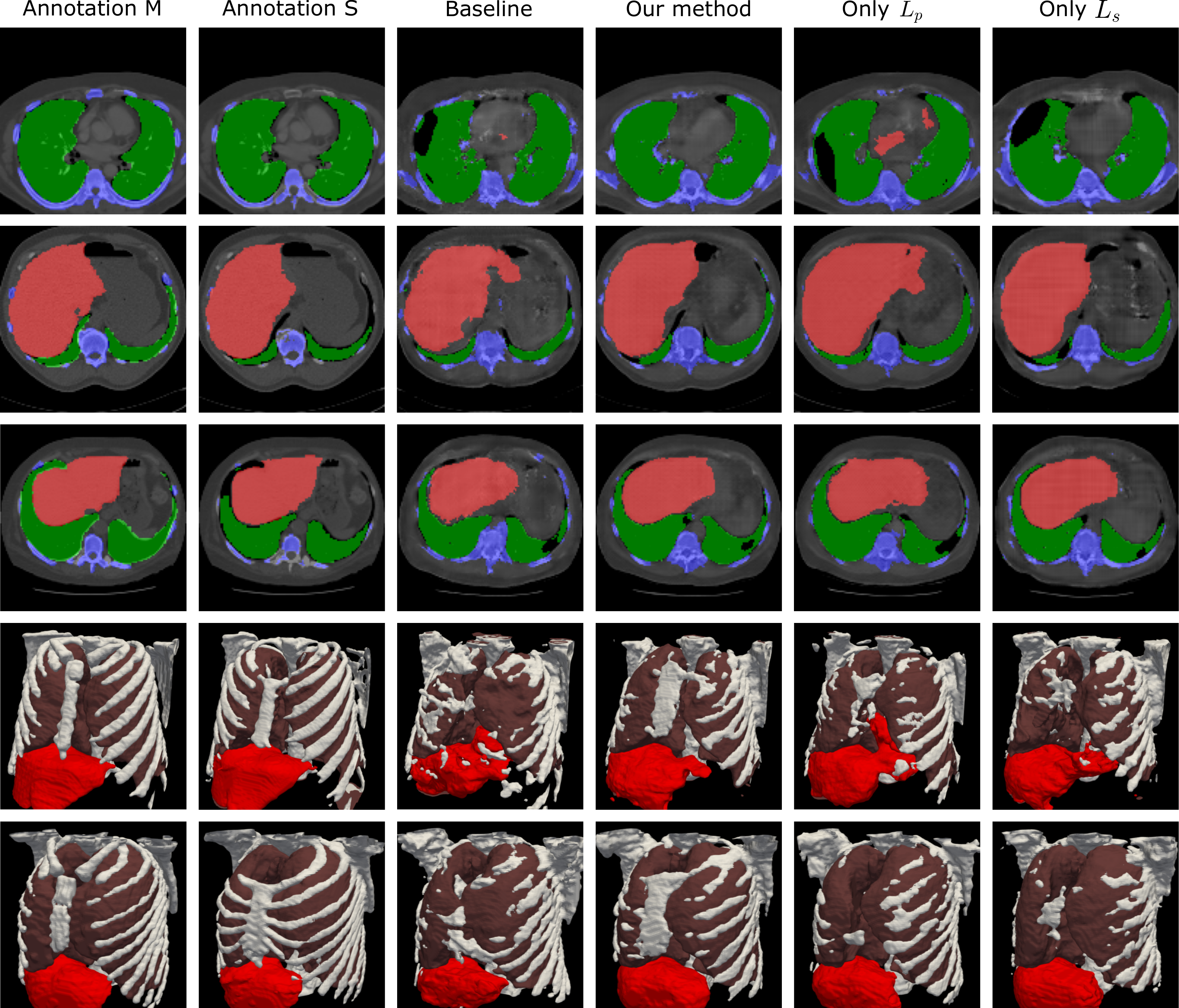}
        \caption{\newstuff{Comparison of the organ segmentation masks generated in different experiments. In the top three rows, the slices of the organ segmentation mask and the CT volumes are shown and in the bottom rows the organ segmentation is shown as mesh visualization.}
        }
        \label{fig_eval_seg}
\end{figure}

% , using the parallel-beam geometry and the fan-beam geometry.
Some exemplary slices of the reconstructed CT volumes are shown in Figure \ref{fig_res_example} and the results of organ segmentation are shown in Figure \ref{fig_eval_seg}. 
% The GAN model by Ying et al. is implemented as the baseline model, in order to illustrate the anatomical enhancement of the proposed loss items.
After evaluating the reconstruction performance, we choose the $\lambda_s=2.0$ and $\lambda_p=0.5$ to present the reconstruction performance of our proposed method.
We demonstrate that our proposed method can improve both the overall image quality and the anatomical structures, in comparison with the baseline method. 
More specifically, \ploss leads to the improved image quality while \sloss can improve the anatomical plausibility of organs and structures. 
The influence of the \ploss and \sloss are further investigated in the ablation experiments, where the either \ploss or \sloss is applied for enhancement with varying $\lambda$s.

% \subsection{Result of the parallel-beam geometry}

% TODO: Update figure with differed lambda

\begin{table}[ht]
    \centering
    \caption{Reconstruction and organ segmentation results of the experiments. $\mathrm{DSC}_{\mathrm{M}}$ indicates the DSC with manual annotation as ground truth and $\mathrm{DSC}_{S}$ indicates the DSC with the segmentation using pre-trained segmentation network as ground truth. The segmentation metrics are aggregated from 20 CT volumes.
    }
    \label{tab_result_pb}
    \begin{tabular}{lcccccccc} \toprule
              & \multicolumn{2}{c}{Baseline} & \multicolumn{2}{c}{Proposed Method} & \multicolumn{2}{c}{Only \ploss} & \multicolumn{2}{c}{Only \sloss} \\ \cmidrule(lr){2-3}\cmidrule(lr){4-5}\cmidrule(lr){6-7}\cmidrule(lr){8-9}
              & avg. & std. & avg. & std.& avg. & std.& avg. & std. \\ \midrule 
              
    PSNR/dB      & 26.21 & 1.02  & 26.49 & 1.29   & 26.53 & 1.03    & 25.52  & 1.19      \\
    SSIM         & 0.62  & 0.06  & 0.64  & 0.05   & 0.64  & 0.05    & 0.60   & 0.06      \\ 
    RMSE/HU       & 201.55 & 24.22 & 196.17 & 31.63 & 195.63 & 33.29 & 219.28 & 33.30 \\ \midrule
    % $\mathrm{DSC}_\mathrm{M}$  & 0.63  & 0.02  & 0.67  &  0.03  & 0.67  &  0.02   & 0.67   & 0.01      \\  
    % $\mathrm{DSC}_\mathrm{S}$ & 0.71  & 0.03  & 0.76  &  0.03  & 0.74  &  0.03   & 0.76   & 0.02      \\ 
    $\mathrm{DSC}_\mathrm{M}$  & 0.63  & 0.02  & 0.71  &  0.03  & 0.69  &  0.02   & 0.70   & 0.01      \\  
    $\mathrm{DSC}_\mathrm{S}$ & 0.71  & 0.03  & 0.76  &  0.03  & 0.74  &  0.03   & 0.76   & 0.02      \\ 
     \bottomrule
         
    \end{tabular}
\end{table}

\begin{table}[ht]
    \centering
    \caption{Organ-wise segmentation results of the proposed method in comparison with the baseline method.
    }
    \label{tab_organ_result}
    \begin{tabular}{lcccccccc} \toprule
              & \multicolumn{4}{c}{Baseline} & \multicolumn{4}{c}{Proposed Method} \\ \cmidrule(lr){2-5}\cmidrule(lr){6-9}
              & Liver & Lung & Bones & Avg. & Liver & Lung & Bones & Avg. \\ \midrule 
    $\mathrm{DSC}_\mathrm{M}$  & 0.74  & 0.81  & 0.35  &  0.63  & 0.82  &  0.86   & 0.45  & 0.71      \\  
    $\mathrm{DSC}_\mathrm{S}$  & 0.75  & 0.84  & 0.57  &  0.71  & 0.81  &  0.86   & 0.63  & 0.76      \\ 
%    $NSD_{GT}$   & 0.51  & 0.02  & 0.54  &  0.02  & 0.55  &  0.09   & \textbf{0.56}   & 0.03      \\
%    $NSD_{GTseg}$& 0.63  & 0.15  & 0.68  &  0.05  & \textbf{0.71}  &  0.02   & 0.69   & 0.02      \\    
     \bottomrule
         
    \end{tabular}
\end{table}

% Mean dice score (DSC) of the three organs is used to evaluate the segmentation performance of the model. Since the LIDC-IDRI dataset contains no paired organ segmentation annotation, 20 CT volumes in the test set are manually annotated with lung, liver and bones, namely annotation $M$. In addition to the manual ground truth, the organ segmentation masks by the pre-trained $S$ are also used to benchmark segmentation performance, as an alternative to the annotation from the same staff who generate the CT-ORG dataset, named as annotation $S$.
% As a support illustration of anatomical plausibility, the bone segmentation is visualized as a 3D mesh. The voxel-wise bone segmentation in the reconstructed CT volumes is firstly converted to surfaces using marching cube algorithm \cite{lorensen1987marching} and then displayed using PyVista (version 0.38.5) \cite{sullivan2019pyvista}.

\subsubsection*{Reconstruction performance} 
% the reconstruction performance of each model are evaluated using the same test data. 

From the exemplary slices in Figure \ref{fig_res_example}. 
The reconstructed CT volumes from the baseline method are 'visual real' but only in terms of the body shape and regions with obvious contrast, such as the boundary of the lungs. 
However, the abdominal organs, for example liver, are not distinctive from the remainder regions. 
Some structural details, like the shape of vertebra, are also lost. 
% The details in the reconstructed CT volumes are lost, for example the shape of the vertebrae in the reconstruction is blurred. 
The proposed method leads to the enhancement of such anatomical structural details and the organ contrast, while keeping the overall image quality.
% Applying the 3D segmentation loss leads to organ-specific enhancement. As the $S$ in our experiments is only trained for lung, liver and bones segmentation, the contrast of such organs in the reconstructed CT volumes are enhanced while there is no enhancement for other organs and structures. 
\newstuff{More specifically, the application of \sloss decreases the image quality. It is expected as \sloss only enhances the organ segmentation rather than the voxel-wise image quality. Regarding the anatomy, \sloss contributes to organ-specific enhancement. As \sloss in our experiments only targets for lung, liver and bones segmentation, the contrast of such organs in the reconstructed CT volumes are enhanced while there is no enhancement for other organs and structures. 
With \ploss the anatomical structures in the CT volumes are enhanced, such as the shape of vertebrae, and the contrast between adjacent anatomical structures, such as the boundary of the fat tissues.
However, such anatomical improvements are barely indicated by the reconstruction metrics. 
Peak-signal-to-noise ratio (PSNR), structural similarity index (SSIM) and root mean squared error (RMSE) in the unit of Hounsfield unit (HU) are selected to evaluate the reconstruction performance. 
Table \ref{tab_result_pb} shows the results of our proposed method in comparison with the baseline method. The proposed method leads to the improvement in the PSNR by 1.0\% and the SSIM by 3.2\%, and RMSE by 2.7\%. With only \sloss, the PSNR is deteriorated by 2.6\%, SSIM by 3.2\% and RMSE by 8.7\%. The best improvement in metrics is obtained with only \ploss, the improvement is by PSNR 1.2\%, SSIM by 3.2\% and RMSE by 2.9\%. From the reconstruction metrics, only \ploss can contribute to the improved image quality.
The results from the ablation study are shown in Figure \ref{fig_lambda_curve}, higher $\lambda_p$ can lead to higher PSNR and SSIM, indicating higher overall image quality. In contrast, higher $\lambda_s$ will not improve the overall image quality. The proposed method also results in higher PSNR and RMSE when $\lambda_s$ and $\lambda_p$ increase, similar to the results with only \ploss.}

\subsubsection*{Organ segmentation in reconstruction}
In additional to the reconstruction metrics, we also evaluate the organ segmentation of the reconstructed volumes for assessment of human anatomy. In our experiments, the segmentation of liver, lung and bones are evaluated, as defined by \sloss. 
% the segmentation of liver, lung and bones, as defined by \sloss, in the reconstructed volumes by the same pre-trained segmentation network is used to assess the anatomical information contained in the reconstructed volume.
Since the reconstruction dataset contains no paired organ segmentation annotation, 20 CT volumes in the test set are manually annotated with such organs, namely annotation $M$. In addition to the manual ground truth, the organ segmentation masks by the pre-trained segmentation network of \sloss are also used to benchmark segmentation performance, named as annotation $S$.
the dice similarity coefficient (DSC) of each organ is then computed. 

\newstuff{Evaluated using the annotation $M$, the proposed method leads to the increase by 12.6\% in average DSC compared with the baseline method, also 9.5\% when only \ploss applied and 11.1\% when only \sloss applied. when the annotation $S$ is used as ground truth, the proposed method leads to the increase by 7.0\% in average DSC compared with baseline method, also 4.2 \% when only \ploss applied and 7.0 \% when only \sloss applied.
In terms of each single organ, as shown in Table \ref{tab_organ_result}, the proposed method improves the $DSC_M$ by 15.1\% for bones, 10.9\% by liver and 6.1\% by lungs. Also the $DSC_S$ is increased by 8.0\% for liver and 10.5\% for bones.}

% The 3D segmentation loss leads to higher increase in DSC when using the annotation $S$ as ground truth, indicating that such enhancement can produce enhanced CT reconstruction preferred by the applied $S$. 

\begin{figure}[ht]
   \centering
   \includegraphics[width=14cm]{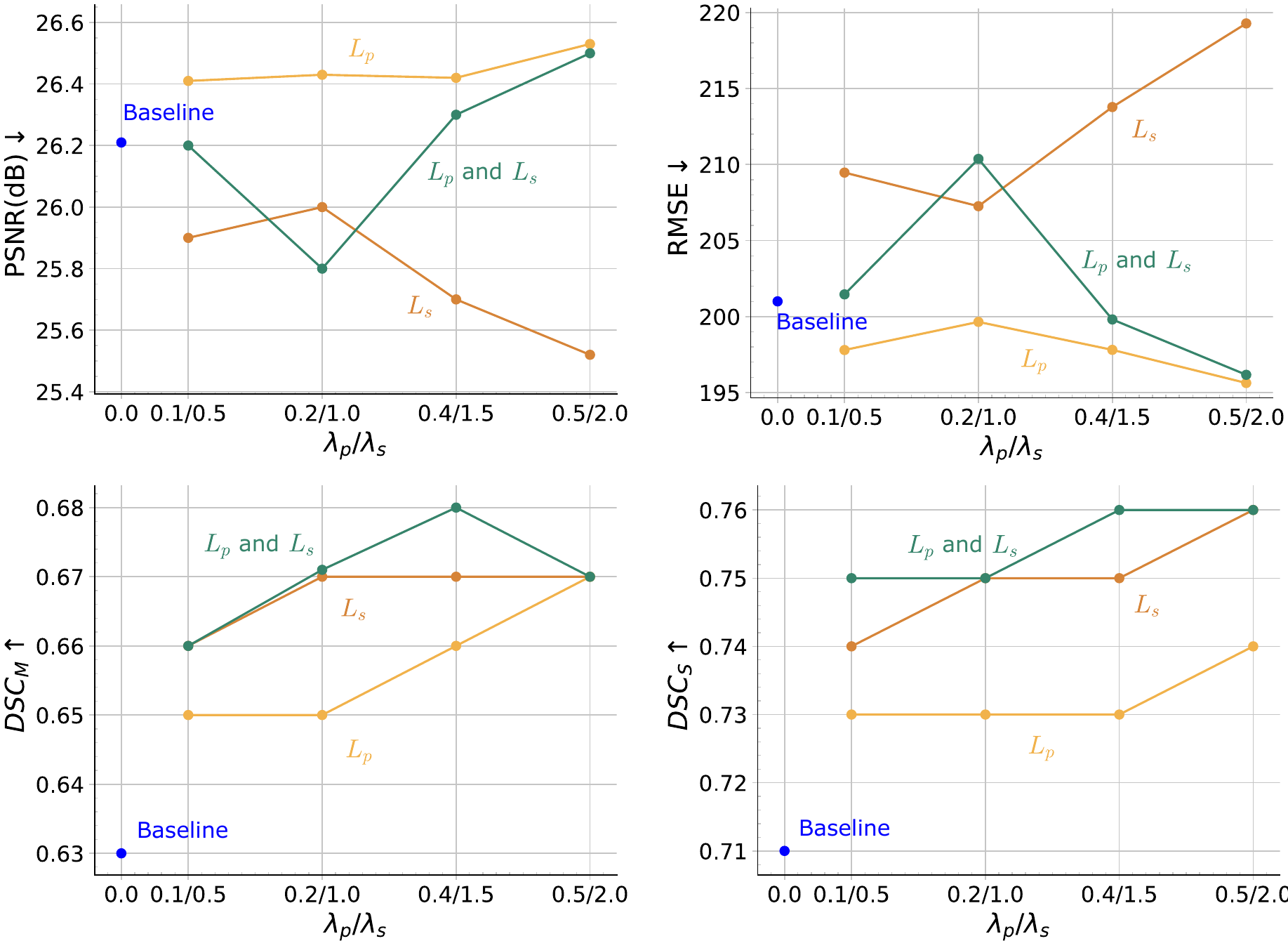}
   \caption{Reconstruction and organ segmentation performance of the proposed method with varying $\lambda$, in comparison with applying \ploss and \sloss independently.
    }  %note label inside caption
   \label{fig_lambda_curve} 
\end{figure}

Some exemplary organ segmentation masks are shown in Figure \ref{fig_eval_seg}. The organ segmentation using the proposed method shows higher anatomical plausibility in terms of the organ and skeleton shape, as shown by the mesh visualization in Figure \ref{fig_eval_seg}.
In comparison, the baseline method and the model with only \ploss contains more outliers and the segmentation of the skeleton is less accurate.
As also shown in Figure \ref{fig_lambda_curve}, higher $\lambda_s$ in general leads to higher average $\mathrm{DSC}_{M}$ and $\mathrm{DSC}_{S}$.

% In general, in our experiments it is found that the application of the 3D segmentation loss can improve the organ segmentation, in terms of average DSC, and the perceptual loss leads to improved image quality, in terms of PSNR and SSIM. Incorporating 3D segmentation loss and the perceptual loss can maintain the improvement in both image quality and organ segmentation. 

% Figure of segmentation examples.
% \begin{figure}[ht]
%     \centering
%         \includegraphics[width=16cm]{Figure_9.pdf}
%         \caption{Comparison of the organ segmentation masks generated in different experiments. In the top three rows, the slices of the organ segmentation mask and the CT volumes are shown and in the bottom row the bone segmentation is shown as mesh visualization.
%         }
%         \label{fig_eval_seg_pb}
% \end{figure}

\section*{Discussion}
Based on the results, the proposed \ploss and \sloss contribute to the enhancement of both the anatomical structures and the overall image quality.
% When applied to a prospective CT risk minimization pipeline, our proposed method bridges the X-ray projections with the reconstructed CT volumes with the segmentation of the OARs, which facilitates the inference of organ dose from only two X-ray projections. 
Such enhancements enable the GANs to reconstruct CT volumes that not only appears correct but also ensures the reliability of the anatomical structures in the reconstructed volumes.
% Thus a more robust reconstruction method for a a prospective CT risk minimization pipeline is provided. 
Consequently, a more robust reconstruction method for a prospective CT risk minimization pipeline is established.
However, the accurate inference of the radiation risk involves more organs as liver, lungs and bones in our research, and aggregating region should be whole human body. It is still difficult to acquire the dataset containing whole body CT scans with all organ segmented for dose estimation, but the research towards larger body region and the segmentation of more diverse organs is our future research of interest. 

\begin{figure}[ht]
\centering
\includegraphics[width=14cm]{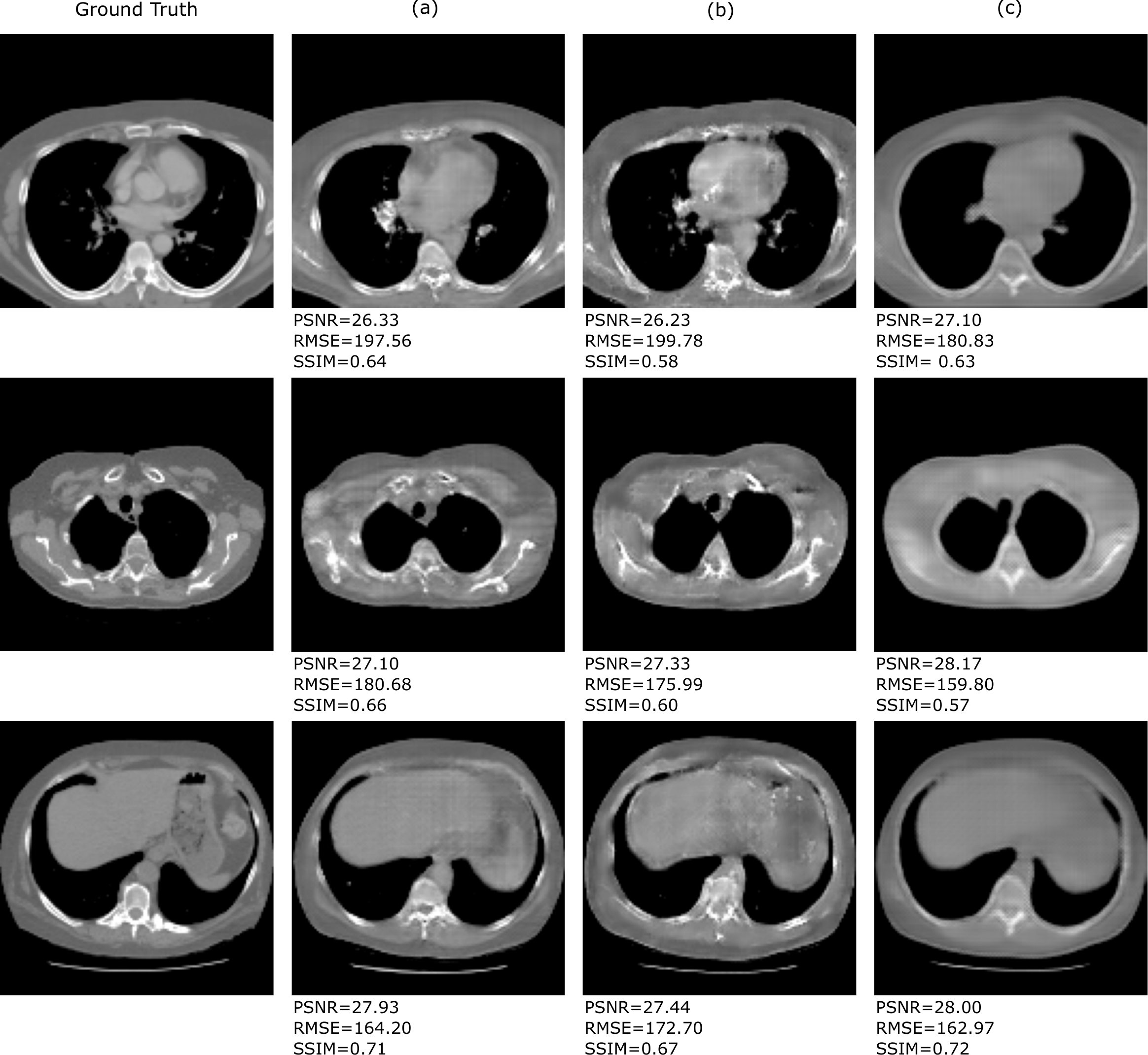}
\caption{Example slices that GANs fails to reconstruct anatomical structures in volumes. Column (a) is our proposed method with enhanced anatomical structures. (b) and (c) illustrate the CT reconstruction with deteriorated anatomical structure but high reconstruction metrics.}
\label{fig_bad}
\end{figure}

\newstuff{Throughout our investigation, we have noted that the reconstructed volumes with enhanced anatomical structures can lead to inferior reconstruction metrics, i.e. PSNR, SSIM and RMSE.
PSNR and RMSE are commonly used for the evaluation of reconstruction algorithms, and SSIM is originally designed for the assessment of digital image quality. 
% Conventional CT reconstruction algorithms make use of full or almost full projections, thus the major factor downgrading the image quality is rarely the missed information.
Different from typical CT reconstruction methods, GAN-based methods depend on training a generator network to reconstruct the volumes from bi-planar projections, so such reconstruction is an ill-posed problem. 
During training the GANs, the network tends to reconstruct the CT volumes with bare or even no anatomical information, while maintaining high reconstruction metrics such as PSNR and SSIM. Some exemplary slices are shown in Figure \ref{fig_bad}. 
% Therefore, in our research the segmentation performance using a pre-trained segmentation network is also applied for assessment, based on the assumption that anatomical plausibility can be well assessed by a network for organ segmentation.
Therefore, in our research we also evaluate the organ segmentation of the CT volumes, based on the assumption that a network that is trained for organ segmentation can effectively evaluate anatomical structures.}

% However, this preliminary study was limited to only a few datasets and a statistically more meaningful evaluation would be required. However, this is out of the scope for this work but could be the topic of future research.

\section*{Methods}

\begin{figure}[ht]
\centering
\includegraphics[width=16cm]{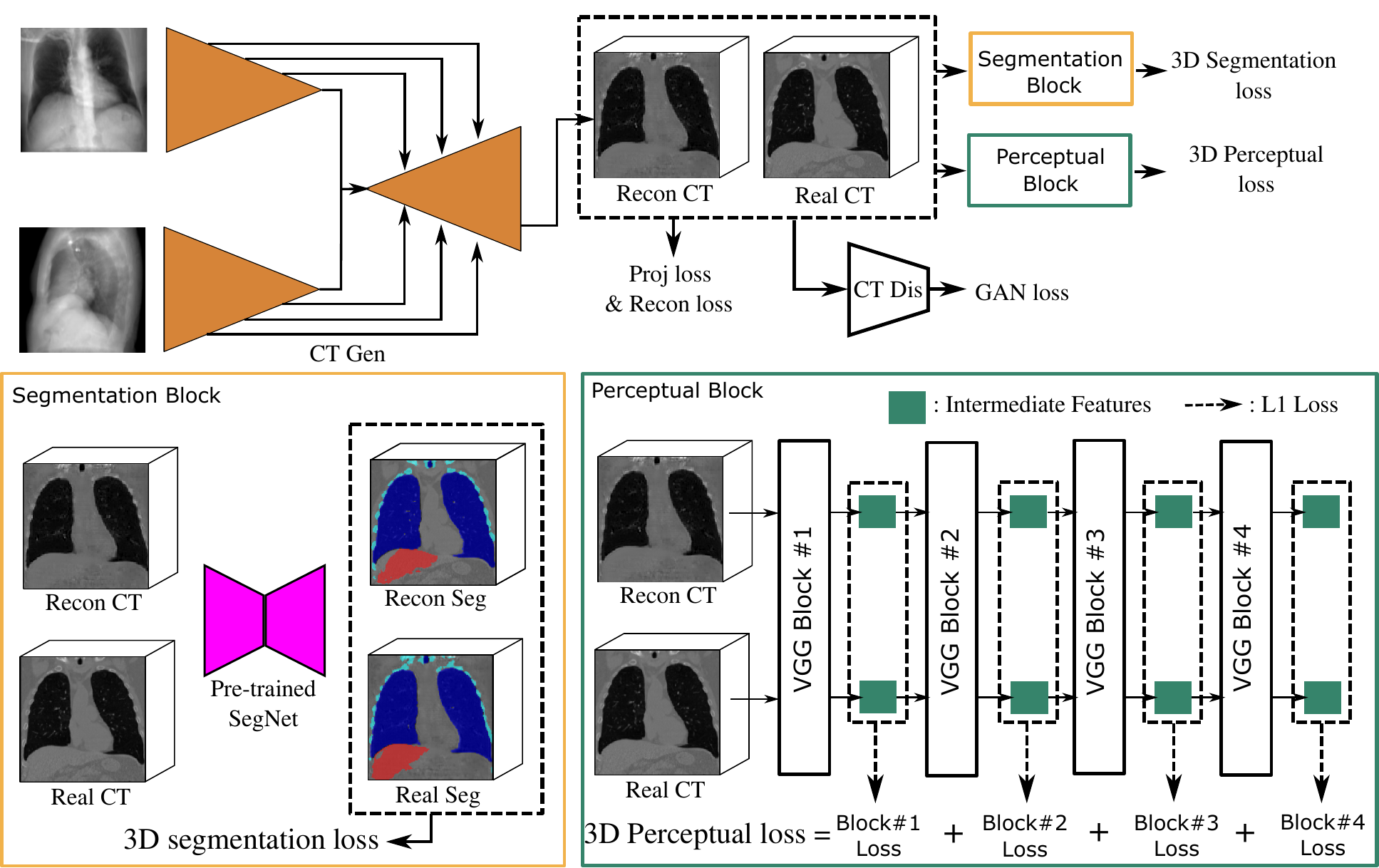}
\caption{The proposed CT reconstruction pipeline with enhancement of anatomical structures. The CT Gen is the CT generator network that reconstructs the 3D CT volume with two 2D X-ray projections as input. The SegNet is a pre-trained segmentation network that segments the target anatomical structures and is frozen during the GAN training. The 3D perceptual loss aggregates the 2D perceptual losses from slices along vertical directions.}  %note label inside caption
\label{fig_pipeline} 
\end{figure}

The pipeline of the proposed model is shown in Figure \ref{fig_pipeline}. A GAN is trained to reconstruct a CT volume from two X-ray projections. On top of the typical generator and discriminator network of GAN \cite{goodfellow2014generative}, two pre-trained networks are included into the training procedure, i.e. a pre-trained segmentation network, namely $\phi_s$, for the enhancement of specific anatomical structures and a pre-trained VGG network for the enhancement of the image quality \cite{Simonyan15}. VGG network is firstly proposed by the visual geometry group (VGG) from the university of Oxford and is a well-known network for image feature extraction in computer vision researches. 

\subsection*{CT reconstruction GAN}

\begin{figure}[ht]
\centering
\includegraphics[width=16cm]{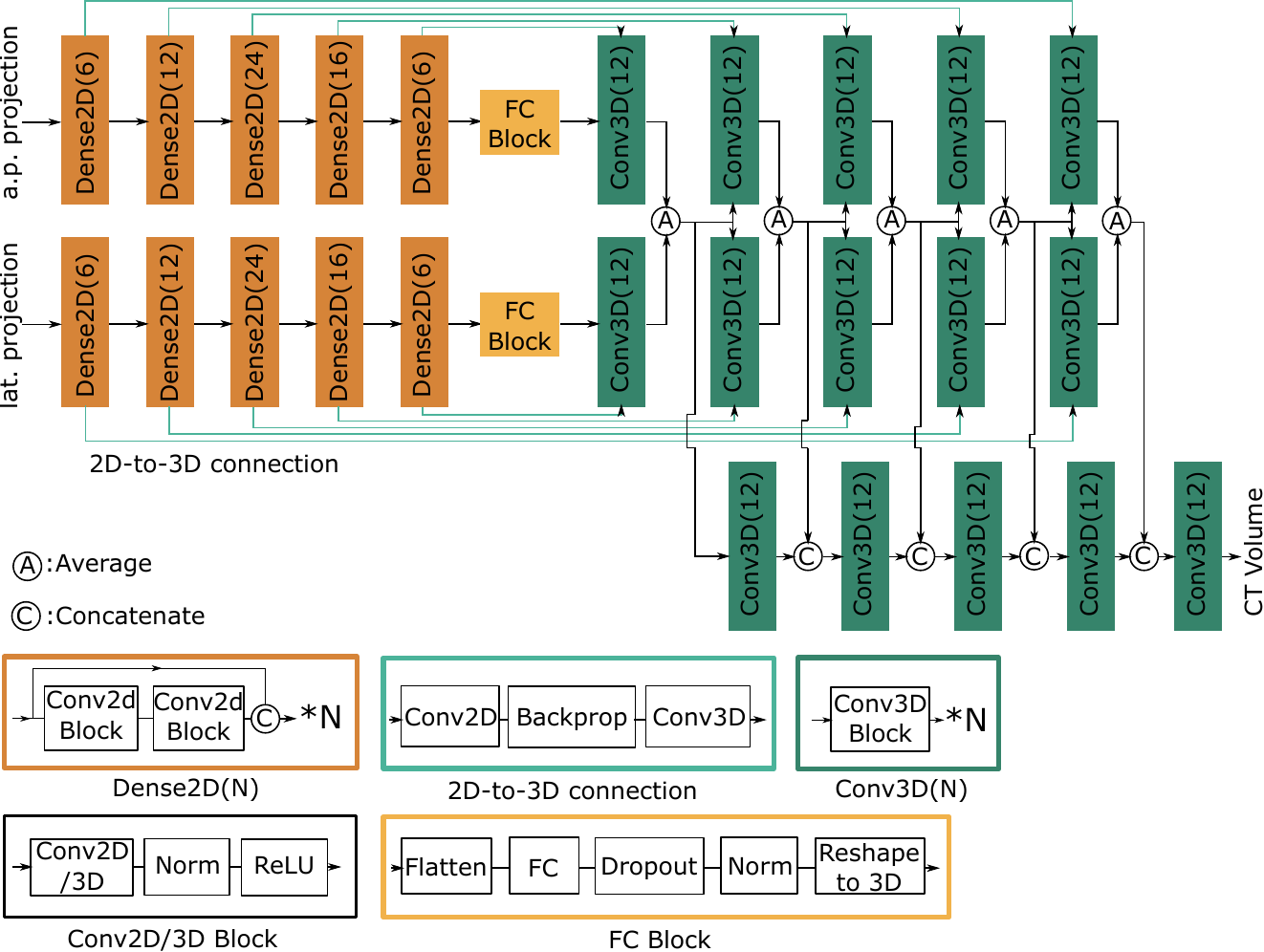}
\caption{Network architecture of the CT generator network that inputs 2D X-ray projections but outputs 3D CT volumes. 
}  %note label inside caption
\label{fig_network_architecture} 
\end{figure}

The training of our proposed model follows the adversarial strategy of GAN. The minmax objective of GAN training in our situation is \cite{goodfellow2014generative}
\begin{equation}
    \underset{G}{\mathrm{min}}\,\underset{D}{\mathrm{max}}\,L_{\mathrm{GAN}} = \mathbb{E}_y[\log D(y)] + \mathbb{E}_x[\log(1-D(G(x))],
\end{equation}
where $x$ indicates the input X-ray projections and $y$ the corresponding CT volume, $G(x;\theta_g)$ and $D(y;\theta_d)$ are the generator and discriminator network. More specifically, the GAN loss is modified according to least squared GAN as two loss functions \cite{mao2017least}

\begin{align}
    & L_{\mathrm{dis}}(x, y) = \frac{1}{2}(\mathbb{E}_{y}\left \|D(y)-1\right \|_2^2+\mathbb{E}_{x}\left \|D(G(x))-0\right \|_2^2), \\
    & L_{\mathrm{gen}}(x) = \mathbb{E}_{x}\left \|D(G(x))-1)\right \|_2^2.
\end{align}
In our model, the discriminator network is implemented as in the work of Phillip et al.\cite{isola2017image}. The generator network encodes the input 2D X-ray projections using two independent pathways based on U-Net \cite{unet} and the encoded features are fused to output the reconstructed CT volume in 3D. The network architecture of the generator network is shown in Figure \ref{fig_network_architecture}. 

% One key step for the generator network is to convert the extracted features from 2D to 3D, and in our architecture such conversion is accomplished using backprojection, as the 2D X-ray projections are obtained by the forward projection of the CT volume. In our work, we model the projection geometry as parallel-beam projection rather than the real fan-beam projection, because parallel-beam projection simplifies the conversion between 2D and 3D intermediate features in the network and can be used to prove the principle of anatomy enhancement. 

One key step for the generator network is to convert the extracted features from 2D to 3D, and in our network such conversion is accomplished using backprojection, as the 2D X-ray projections are obtained by the forward projection of the CT volumes. 
% In our work, both parallel-beam and fan-beam geometries are researched. 
fan-beam geometry is implemented in our research.
% As shown in Fig. \ref{fig_fanbeam}, we design a fan-beam operator to model the fan-beam backprojection, in particular the geometry of a Siemens Somatom Force scanner. 
The backprojection propagates the 2D feature maps to 3D and is implemented as a matrix multiplication,

\begin{equation}
    \hat{Z}_{3D} = T \cdot \hat{Z}_{2D},
\end{equation}
where $\hat{Z}$ is the flattened 2D or 3D intermediate feature maps and $T$ is a pre-defined transformation matrix depending on the fan-beam geometry. 
% For parallel-beam geometry, $T$ is a parallel-beam backprojector using orthogonal projection. 
In this work $T$ is given by a pixel-driven fan-beam backprojector based on the geometry of a Siemens Somatom Force scanner.

% \begin{figure}[ht]
% \includegraphics[width=11cm]{Figure_5.pdf}
% \caption{Comparison of the parallel-beam operator (up) and the geometry-based fan-beam operator (down). In our network, such operator is implemented to simulate the propagation of 2D feature maps to 3D as backprojection.
% }  %note label inside caption
% \label{fig_fanbeam} 
% \end{figure}

\subsection*{3D segmentation loss} \label{sec_segloss}
% We first propose the \sloss for the enhancement of specific anatomical structures.
\sloss is first proposed for the enhancement of specific anatomical structures.
The correct location, shape and size of the OARs in the reconstructed CT volumes are crucial for dose estimation and organ segmentation, but such anatomical content cannot be explicitly leveraged using typical image generation models, such as GANs. In order to include also the organ segmentation into the training of the GAN, a dataset of CT volumes with the segmentation ground truth is required. However, the voxel-wise annotation of the OARs is very expensive and cannot be easily obtained for large-scale dataset for training a GAN, while the segmentation datasets are mostly not sufficient in the number of images for training GANs.

In our model, a pre-trained \segnet is leveraged to enhance the anatomical content that are missing in the reconstruction dataset. The \segnet is trained on an auxiliary dataset that contains the segmentation of the OARs in CT volumes. Such a \segnet is then applied into the training of the GAN and the enhancement of the anatomical structures is thus explicitly refactored to the optimization of OARs segmentation in the reconstructed CT based on the pre-trained \segnet.  Such regularization is implemented as a loss item 

\begin{align}
    L_s(y'_m, y_m)&=1-2\frac{\sum y_my'_m}{\sum y_m + \sum y'_m},\\
\end{align}
where $y_m=\phi_s(y)$ and $y'_m=\phi_s(G(x))$ are the organ segmentation mask of $y$ and $G(x)$ using \segnet. From the equation, \sloss will depend on the target organ of the \segnet, so the enhancement of anatomical structures can be flexible to specific organs. Since the segmentation ground truth of the reconstruction dataset is missing, the \segnet will not be optimized during GAN training. After the training of the GAN, the \segnet can be further used to provide organ segmentation. During the inference, the model outputs $G(x)$ as reconstructed CT volume and $\phi_s(G(x))$ as the corresponding organ segmentation map with $x$ the input X-ray projections.

\subsection*{3D perceptual loss} \label{sec_ploss}
Perceptual loss is first proposed in the field of computer vision for feed-forward image transformation tasks \cite{johnson2016perceptual}. Unlike typical loss functions, perceptual loss relies on a pre-trained classification network as feature extractor and backpropagates the loss using the extracted features from the source and the target images. Apart from natural image researches, perceptual loss is also applied in medical image processing researches, such as the denoising of CT images \cite{yang2018low}. It is shown that the network pre-trained on natural images can also work as a good feature extractor for medical images. In our model, we adopted the VGG-16 network pre-trained on the ImageNet dataset as the feature extractor, deployed by the torchvision toolkit (version 0.15.2)\cite{Daniel2022torchvision, deng2009imagenet}. The original VGG-16 network contains five convolutional blocks to extract image features in different scales. The aggregation of L1 loss of intermediate features from the ground truth and the reconstructed CT volumes leads to the 3D perceptual loss, as shown in Figure \ref{fig_pipeline}. The 3D perceptual loss used in the model training is 

\begin{equation}
    L_p(y', y)=\mathbb{E}_{y,y'}\left \| \phi_p(y)-\phi_p(y') \right \|_2^2,
\end{equation}
where the $\phi_p()$ is the intermediate features and only the first four VGG levels are used to aggregate the 3D perceptual loss. Note that the pre-trained VGG network only inputs 2D images, so the ground truth and the reconstructed CT volumes are sliced along the vertical direction and the loss of all 2D slices are aggregrated.

\subsection*{Overall loss function}
In addition to the previously mentioned loss functions, the voxel-based $L_{\mathrm{r}}$ and pixel-based $L_{\mathrm{proj}}$ are applied for the consistency of the input X-ray projections and the reconstructed CT volume, which is implemented as

\begin{align}
    L_{\mathrm{r}}(y,y')&=\mathbb{E}_{y,y'}\left \| y-y'\right \|_2^2, \\
    L_{\mathrm{proj}}(y,y')&=\frac{1}{2}\mathbb{E}_{y,y'}[\left \| P_{\mathrm{a.p.}}(y)-P_{\mathrm{a.p.}}(y')\right \|_2^2 + \left \| P_{\mathrm{lat.}}(y)-P_{\mathrm{lat.}}(y')\right \|_2^2],
\end{align}
where $P_{\mathrm{a.p.}}$ and $P_{\mathrm{lat}}$ project the CT volume each in a.p. and lat. direction. $L_r$ will lead to the CT reconstruction to be correct and $L_{proj}$ will lead to the projections to be correct. 
Then with the proposed \sloss and \ploss, the overall loss function for training the CT reconstruction GAN is weighted to balance the voxel-wise features and anatomical contents.
The overall loss function aggregates as 

\begin{align}
    L_D & = L_{\mathrm{dis}} \\
    L_G & = \lambda_{\mathrm{gen}}L_{\mathrm{gen}} + \lambda_\mathrm{r}L_{\mathrm{r}} + \lambda_\mathrm{proj}L_{\mathrm{proj}} + \lambda_\mathrm{s}L_{\mathrm{s}} + \lambda_{\mathrm{p}}L_{\mathrm{p}}
\end{align}
where $\lambda$s are configurable hyper-parameters, and in the experiments we will illustrate how the CT reconstruction is enhanced by the proposed model.

\subsection*{Datasets}
% Here 2 or more datasets have been used. Explain it.

For the training of the GAN and the pre-training of the \segnet, two datasets are used in our experiment for the proof of the principle, i.e. a reconstruction dataset and a segmentation dataset. 
For training the GAN, the CT volumes from the lung image database consortium and image database resource initiative (LIDC-IDRI) are used as the reconstruction dataset \cite{Armato2011-lo}. The LIDC-IDRI dataset consists of 1016 chest CT volumes with pixel size ranging from 0.46\,mm to 0.98\,mm in the transverse plane and from 0.6\,mm to 5.0\,mm in vertical direction. 
% Besides the chest region, we also used the AMOS dataset to investigate our proposed model on abdominal region \cite{ji2022amos}. The AMOS dataset consists of 300 abdominal CT volumes.

\begin{figure}[ht]
\centering
\includegraphics[width=10cm]{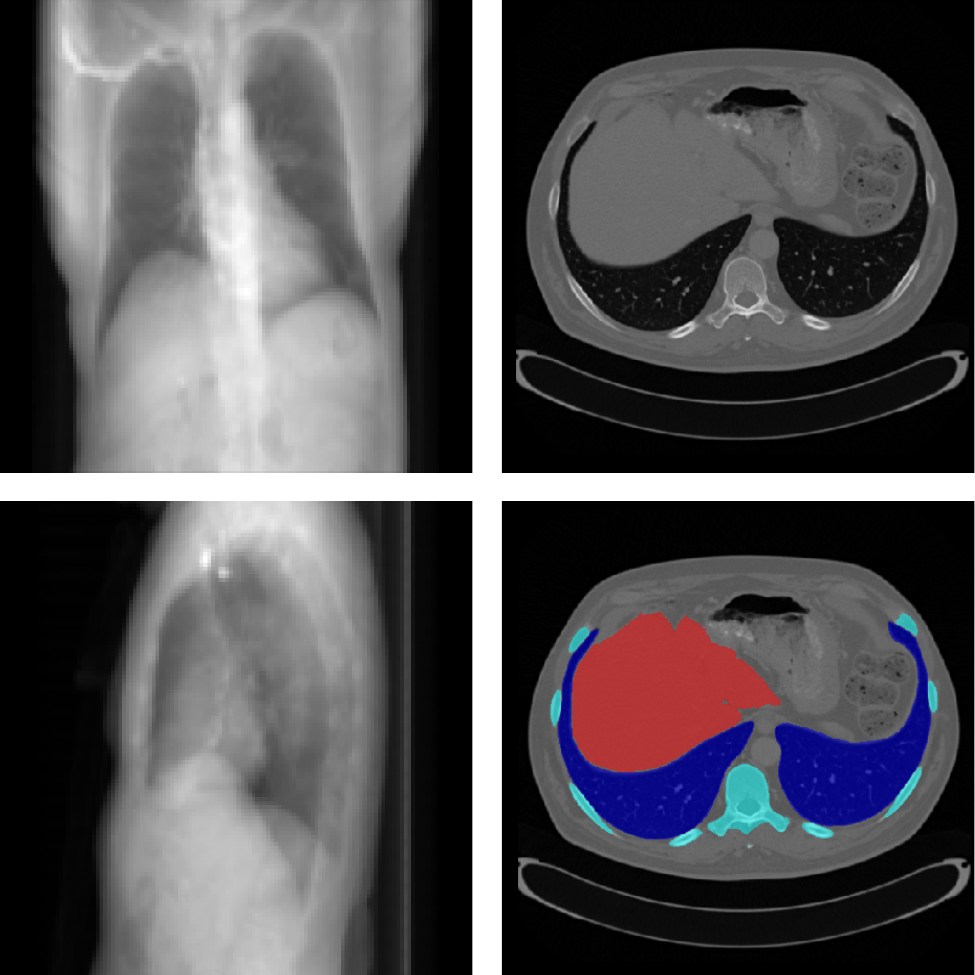}
\caption{Example simulated X-ray projections from the LIDC-IDRI dataset and fan-beam geometry (left) and one slice from the CT volume with organ segmentation from CT-ORG dataset (right).
}  %note label inside caption
\label{fig_example_data} 
\end{figure}

For the pre-training of the \segnet, we select a public dataset of CT volumes with voxel-wise annotation of abdominal organs, namely CT-ORG dataset \cite{Rister2020-gw}. The CT-ORG dataset consists of 140 throat-abdominal CT scans with annotated lungs, bones, liver, bladder and brain, with voxel size ranging from 0.56\,mm to 1.0\,mm in vertical direction. Because the reconstruction dataset covers only chest region, the annotations of lungs, liver and bone in the CT-ORG dataset are used in the following experiments. Some samples from the datasets are shown in Figure \ref{fig_example_data}. All CT volumes in the LIDC-IDRI dataset and the CT-ORG dataset are resampled to a uniform voxel size of 1\,mm by 1\,mm by 1\,mm to ensure the consistency during model training. 812 CT images from the LIDC-IDRI dataset are used during the model training and 214 images for test. For training the \segnet, 112 CT scans are used for training and 28 images for test. 
% For investigation of the abdominal region, 240 images from the AMOS dataset are used for model training and 60 images for test.
% The X-ray projections are simulated in a way to mimic the CT parallel-beam projection, by averaging the voxel values along the a.p. and lat. directions. For the GAN training, both X-ray projections and the CT volumes are normalized to 0.0 to 1.0 using the same parameters.
The X-ray projections are simulated in a way to mimic the fan-beam CT forward projection, by using the aforementioned scanner geometry. The CT volumes are first resampled to voxel size of 2.5\,mm in each direction and then clipped to the uniform volume/image size of 128. The resolution of the input X-ray projections is also 128. For the GAN training, both X-ray projections and the CT volumes are normalized to 0.0 to 1.0 using the same parameters.

\subsection*{Experiments}\label{sec_experiment}

% In our experiments, we initially test our proposed model using the parallel-beam geometry and then the fan-beam geometry. For each geometry, the enhancement of lung, liver and bones in the CT reconstruction is tested by using 3D segmentation loss and 3D perceptual loss.
% % In our experiments, we tested our proposed model and the enhancement of lung, liver and bones in the CT reconstruction using both the 3D segmentation loss and the 3D perceptual loss. 
% The proposed methods are evaluated in comparison with the baseline method, which contains only the GAN loss, the reconstruction loss and the projection loss. For a closer look into how differently the 3D segmentation loss and the 3D perceptual loss influence on the CT reconstruction, the experiments when applying either the 3D segmentation loss or the 3D perceptual loss are also done. In addition, ablation experiments are done where $\lambda_{\mathrm{s}}$ and $\lambda_{\mathrm{pt}}$ are varied within the range of 0.1 to 2.0. 

In all experiments, the GAN is trained for 100 epochs. Adam optimizer is used with learning rate of $2\cdot10^{-4}$. The weights of the GAN loss, the reconstruction loss and the projection loss, namely $\lambda_{\mathrm{gen}}$, $\lambda_{\mathrm{r}}$ and $\lambda_{\mathrm{proj}}$, are fixed across all experiments, i.e. \lambdarm{gen}=0.1, \lambdarm{r}=10 and \lambdarm{proj}=10. In all experiments, \segnet is implemented as an vanilla 3D U-Net and trained on the CT-ORG dataset for 200 epochs. Dice loss is used as loss function and Adam is used as the optimizer with learning rate of $5\cdot10^{-4}$. All model training is carried out on one Nvidia A100 GPU with 40GB memory.
For the fan-beam operator, we model the real scanner parameters with the source-to-detector distance (SDD) as 1085.6\,mm, the source-to-isocenter distance (SID) as 595\,mm and the number of rays within the fan to be 920.

% How the evaluation is done. 
% After training, the reconstruction performance of each model are evaluated using the same test data. Peak-signal-to-noise ratio (PSNR), structural similarity index (SSIM) and root mean squared error (RMSE) in the unit of Hounsfield unit (HU) are selected to evaluate the reconstruction performance against the ground truth CT volume. 
% Mean dice score (DSC) of the three organs is used to evaluate the segmentation performance of the model. Since the LIDC-IDRI dataset contains no paired organ segmentation annotation, 20 CT volumes in the test set are manually annotated with lung, liver and bones, namely annotation $M$. In addition to the manual ground truth, the organ segmentation masks by the pre-trained $S$ are also used to benchmark segmentation performance, as an alternative to the annotation from the same staff who generate the CT-ORG dataset, named as annotation $S$.
% As a support illustration of anatomical plausibility, the bone segmentation is visualized as a 3D mesh. The voxel-wise bone segmentation in the reconstructed CT volumes is firstly converted to surfaces using marching cube algorithm \cite{lorensen1987marching} and then displayed using PyVista (version 0.38.5) \cite{sullivan2019pyvista}.

\bibliography{sample}

\section*{Acknowledgements}

This work was supported by the Deutsche Forschungsgemeinschaft (DFG) under grant KA 1678/24, LE 2763/3 and MA 4898/15. 

\section*{Author contributions statement}
Must include all authors, identified by initials, for example:
C.L., L.K.,Y.H., M.K. and A.M. conceived the main idea, C.L. performed the experiments and the evaluation, C.L. wrote the main part of the manuscript, L.K. and E.B offered support in the field of medical physics, M.K.,A.M provided expertise through intense discussion. All authors reviewed the manuscript. 

\section*{Additional information}
% To include, in this order: \textbf{Accession codes} (where applicable); \textbf{Competing interests} (mandatory statement). 

% The corresponding author is responsible for submitting a \href{http://www.nature.com/srep/policies/index.html#competing}{competing interests statement} on behalf of all authors of the paper. This statement must be included in the submitted article file.

% \begin{figure}[ht]
% \centering
% \includegraphics[width=\linewidth]{stream}
% \caption{Legend (350 words max). Example legend text.}
% \label{fig:stream}
% \end{figure}

% \begin{table}[ht]
% \centering
% \begin{tabular}{|l|l|l|}
% \hline
% Condition & n & p \\
% \hline
% A & 5 & 0.1 \\
% \hline
% B & 10 & 0.01 \\
% \hline
% \end{tabular}
% \caption{\label{tab:example}Legend (350 words max). Example legend text.}
% \end{table}

% Figures and tables can be referenced in LaTeX using the ref command, e.g. Figure \ref{fig:stream} and Table \ref{tab:example}.

\end{document}